# Resource Allocation in Mobile WiMAX Network: An Optimal Approach

**Rakesh Kumar Jha,\* , Upena D Dalal and Vishal Wankhade A**
 SVNIT / Surat, Gujarat, 395007

E-Mails: jharakesh.45@gmail.com; upena_dala@yahoo.com, wankhedeva@gmail.com
\* Electronics and Communication Engineering Department; Tel.: +91-9328869707

**Abstract:** In the last few years there has been significant growth in the area of wireless communication. IEEE 802.16/WiMAX is the network which is designed for providing high speed wide area broadband wireless access; WiMAX is an emerging wireless technology for creating multi-hop Mesh network. Future generation networks will be characterized by variable and high data rates, Quality of Services (QoS), seamless mobility both within a network and between networks of different technologies and service providers. A technology is developed to accomplish these necessities is regular by IEEE, is 802.16, also called as WiMAX (Worldwide Interoperability for Microwave Access). This architecture aims to apply Long range connectivity, High data rates, High security, Low power utilization and Excellent Quality of Services and squat deployment costs to a wireless access technology on a metropolitan level. In this paper we have observed the performance analysis of location based resource allocation for WiMAX and WLAN-WiMAX client and in second phase we observed the rate-adaptive algorithms. We know that base station (BS) is observed the ranging first for all subscribers then established the link between them and in final phase they will allocate the resource with Subcarriers allocation according to the demand (UL) i.e. video, voice and data application. We propose linear approach, Active-Set optimization and Genetic Algorithm for Resource Allocation in downlink Mobile WiMAX networks. Purpose of proposed algorithms is to optimize total throughput. Simulation results show that Genetic Algorithm and Active-Set algorithm performs better than previous methods in terms of higher capacities but GA have high complexity then active set.

## 1. Introduction

Our proposed work will play a very imperative role in NGN network. Network deployment in any location is first and foremost challenging task. LBRRA has been prearranged suggestion about network performance with respect to orientation of location, population density and user demand. It will assist how you will overcome the problem and how you will planned your network so that performance of deployed network will swell with Cross layer architectures, Power distribution, AMC.OPNET Modeler have features to deployed exact location on MAP with desired terrain .In our case study location has been dispensed for clients and BS in WiMAX and WiMAX-WLAN interface network are Surat, Bardoli, and Navsari (Gujarat, and India). In first case we observed the performance analysis with WiMAX clients. In second case study we observed the performance analysis for same location with WLAN-WiMAX interface scenario (Future scenario).In these condition clients are simple Wireless LAN (Wi-Fi) Nodes getting access from Access Points (AP), these Access Points are connected with WiMAX to WLAN converters. In second phase we have observed LBRRA for the WiMAX network, discuss its Architecture along with the brief explanation of its Physical Layer and MAC (Media Access Control) Layer. The WiMAX network is implemented with the help of OPNET Modeler Networking tool, the performance analysis of the network model is done along with resource allocation i.e. the Base Station (BS) allocates its desired resources by setting the modulation schemes on Subscriber Stations (SS) depending on its distance from Base Station (BS), also to check the total capacity of the network by discussing the admitted and rejected connections to the Base Station (BS), these admitted and rejected connections are shown on the basis of QoS (Quality of Service) Scheduling Services provided by Base Station (BS).In last phase we had compared the performance analysis with AMC. Our proposed algorithm will help o reduce this rejection with optimum approach.**Application Config are shown in Table 1. and Table 2.**

Table 1. WiMAX Configuration Node Parameters

| Service Class Name | Scheduling Type | Maximum Sustained Traffic Rate (bps) | Minimum Reserved Traffic Rate(bps) | Maximum Latency (milliseconds) |
|---|---|---|---|---|
| Gold | UGS | 5 Mbps | 1 Mbps | 30 |
| Silver | rtps | 5Mbps | 64000 | 30 |
| Bronze | Best Effort | 0.5 Mbps | 32000 | 30 |

**Table 2.** Application Configuration Nodes Parameters

| Server (Location) | Application |
|---|---|
| **Hyderabad** | File Transfer |
| **Kolkata** | Video Conferencing |
| **Delhi** | Voice over IP Call |
| **Client(Location)** | Surat,Gujarat,India (With different Demand) |

In case of WLAN – WiMAX interface all the scenarios architecture and application is same only AMC will change. Scenario 1: Optimization with 16 QAM, Scenario 2: Optimization with 64 QAM and in Scenario 3: Optimization with QPSK. In AMC Profile Set Definition there are two types of profile sets, Uplink (UL) Profile Set and Downlink (DL) Profile Sets, each profile sets consists of two rows, Row 0 and Row 1.Minimum entry threshold and mandatory exit threshold with respect to different modulation and coding rates has been defined. The OFDMA parameters for simulation are given in Table3.

**Table 3.** OFDMA PHY Layer Profile

| Frame Duration (ms) | 5 ms |
|---|---|
| Symbol Duration (ms) | 100.8 ms |
| Number of Subcarriers | 2048 |
| Frame Preamble (symbol) | 1 symbol |
| TTG (us) | 106 us |
| RTG (us) | 60 us |
| Duplexing Technique | TDD |
| Base Frequency (GHz) | 5.8 GHz |
| Bandwidth (MHz) | 20 MHz |

## 2. Linear Approach for Resource

In this method rough proportionality assigned to different subscribers while keeping in mind all subscribers get equal resources at the end. This approach is best when all subscribers using same Quality of Service (QoS).

*2.1 Flowchart of Linear Approach*

The set of total power assigned for each user k, denoted as $P_k$ for $K \geq k \geq 1$, can be solved using Lagrange multiplier techniques.

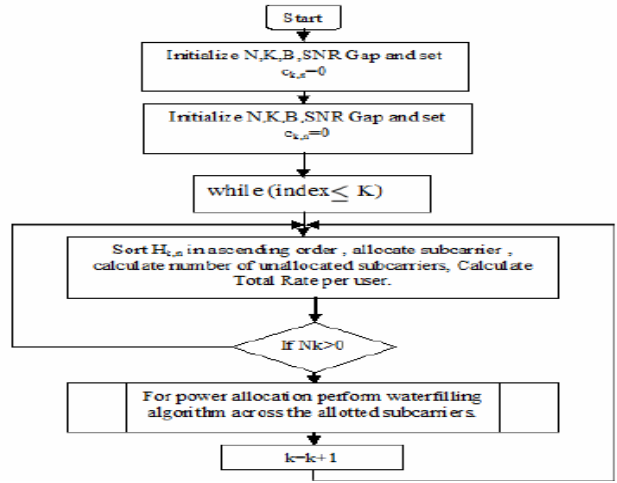

Figure 1. Flowchart for Linear approach

**Experiment 1:**

In this experiment we consider only two subscribers are online, with total number of subcarriers 4 and total power available is 10W. Proportionality assigned to both the subscribers is 75% and 25% respectively for first and second subscriber. Hence first user got 3 subcarriers and second got only 1 subcarrier. E.g. we assumed that effective Subchannels signal to noise ratio vector is a set of {10, 8, 9, and 7} four elements, using these values we can form following table. Finally subscriber one allotted more power i.e. 7.66W as compared to subscriber 2 i.e. 2.34W, hence 10W of power utilized amongst two subscribers. After calculation of total power per subscriber. We can easily calculate data rate.

**Table 4.** Power Allotted And Data Rates for Subscribers 1 and 2 Using Linear Approach Active-set Algorithm for Resource Allocation

| $H_{11}$ | $H_{12}$ | $H_{13}$ | $H_{14}$ |
|---|---|---|---|
| 10 | 8 | 9 | 7 |
| $V_1$ | $V_2$ | $W_1$ | $W_2$ |
| -0.068 | -0.011 | 0.83 | 0.9 |
| $A_{2,2}$ | $B_1$ | $B_2$ | |
| -2.93 | 0 | 0.1632 | |
| $P_1$ (Power allotted for Subscriber1) | | | $P_2$ (Power allotted Subscriber2) |
| $P_{1,1}$ | $P_{1,2}$ | $P_{1,3}$ | $P_{21}$ |
| 2.58 | 2.55 | 2.53 | 2.34 |
| Total power allotted to subscriber 1 | | | Total power allotted to subscriber 2 |
| 7.66 | | | 2.34 |
| Subscriber | | | Data Rate (bits/sec) |
| 1 | | | 13.39008 |
| 2 | | | 4.46336 |

(2) Simulation Experiment 2 (A):

Experiment is done on two subscribers only with 50% proportionality for each user; total subcarriers available are 8, hence 4 subcarriers allotted to each user. We applied both Linear and Active-set algorithm and calculated capacity achieved using both the algorithms.

**Table 5**. Power Allotted and Data Rates for Subscriber 1 and 2 Using an Ative-Set Optimization Algorithm for Experiment 2(a)

| Subcarrier Allocation Matrix ($c_{k,n}$) | User1 | 1 | 1 | 0 | 0 | 0 | 1 | 0 | 1 |
|---|---|---|---|---|---|---|---|---|---|
| | User2 | 0 | 0 | 1 | 1 | 1 | 0 | 1 | 0 |
| $H_{k,n}$ Matrix | User1 | 189 | 265 | 0 | 0 | 0 | 46 | 0 | 87 |
| | User2 | 0 | 0 | 301 | 363 | 288 | 0 | 230 | 0 |
| **Power Allocation by** | | | | | | | | | |
| **Linear** | | | | | **Active-Set** | | | | |
| User1 | | User2 | | | User1 | | User2 | | |
| 0.2929 | | 0.7071 | | | 0.5 | | 0.5 | | |
| **Capacity Achieved in bits/sec/Hz by** | | | | | | | | | |
| **Linear** | | | | | **Active-Set** | | | | |
| 4.65 | | | | | 4.85 | | | | |

(3) Experiment 2 (B):

In this Experiment we added two more subscribers on the network and evaluate the performance of both the algorithms.

**Table 6.** Shows Performance for 4 Subscribers.

| Subcarrier Allocation Matrix ($c_{k,n}$) | User1 | 1 | 0 | 0 | 0 | 0 | 0 | 0 | 1 |
|---|---|---|---|---|---|---|---|---|---|
| | User2 | 0 | 1 | 1 | 0 | 0 | 0 | 0 | 0 |
| | User3 | 0 | 0 | 0 | 0 | 0 | 1 | 1 | 0 |
| | User4 | 0 | 0 | 0 | 1 | 1 | 0 | 0 | 0 |
| $H_{k,n}$ Matrix | User1 | 189 | 265 | 0 | 0 | 0 | 46 | 0 | 87 |
| | User2 | 0 | 0 | 301 | 363 | 288 | 0 | 230 | 0 |
| **Power Allocation by** | | | | | | | | | |
| **Linear** | | | | **Active-Set** | | | | | |
| User1 | User2 | User3 | User 4 | User1 | User2 | User3 | User4 | | |
| 0.356 | 0.382 | 0.1903 | 0.071 | 0.25 | 0.25 | 0.25 | 0.25 | | |
| **Capacity Achieved in bits/sec/Hz by** | | | | | | | | | |
| **Linear** | | | | **Active-Set** | | | | | |
| User1 & 2 | | User3 & 4 | | User1 & 2 | | User3 & 4 | | | |
| 4.2463 | | 3.3577 | | 4.4274 | | 3.839 | | | |

*2.2. Genetic Algorithm for Resource Allocation*

Genetic algorithm (GA) uses three operators, selection, crossover and mutation to direct the population towards convergence at the global optimum. Naturally these initial

*Main Routine for Genetic Algorithm*
*{*

*Initialize population; Evaluate population;*
*While (Termination Criteria Not Satisfied)*
*{*
*Select parents for reproduction;*
*Perform recombination and mutation;*
*Evaluate population;*
*}*
*}*

(1) Experiment 3:

Experiment is done on two subscribers only with 50% Proportionality for each user; total subcarriers available are 8, hence 4 subcarriers allotted to each user. We used same inputs which were given to Active-Set algorithm. We noted down following results for capacity for the two subscribers for 8 iterations or generations.

**Table 7.** Capacity Achieved using GA for two subscribers

| Iteration | Capacity | |
|---|---|---|
| | User 1 | User 2 |
| 1 | 5.0254 | 2.9410 |
| 2 | 3.0452 | 4.3337 |
| 3 | 4.3129 | 4.7474 |
| 4 | 5.6984 | 4.3495 |
| 5 | 6.0764 | 5.6382 |
| 6 | 6.2318 | 6.0812 |
| 7 | 6.2412 | 6.1012 |
| 8 | 6.3098 | 6.1113 |

**3. Result and Analysis**

*3.1. WLAN-WiMAX*

The performance analysis has specified proposal about how radio recourse is allocated in particular location with population density, city orientation and user demand. Tabular results are given comparison for different location based resource allocations.

**Table 8.** Our proposed work result based on LBRRA

| Static | Value |
|---|---|
| Total Capacity (Msps) | 12.710400 (In All Location) |
| Total Uplink Capacity | 2.630400 |
| Total Downlink Capacity | 10.080000 |
| Admitted Capacity | 12.642400 (Surat), 12.613600(Bardoli) and 12.662600 (Navsari) |
| Number Of Admitted Connections | 62 (Surat), 44 (Bardoli) and 63(Navsari) |
| Number Of rejected Connections | 19, 10 and 18 Respectively |
| No of clients | 12, 6 and 9 Respectively |
| Performance Parameters allied with Optimization | With AMC, Cross Layer Optimization and Power optimization |

Our results are replicate that only user density is not prime task for network deployment rather than location. They are also munificent reflection about B.W admitted rejection so that we can improve performance with optimization. One of the optimization has done on the basis of AMC in WLAN-WiMAX network.

In case of WLAN-WiMAX interface scenario compared the result on the basis of admitted connection, rejection connection, Request Bandwidth (RW) and Admitted Bandwidth with AMC is given in Table 9.

**Table 9.** Result based on AMC.

| AMC (Adaptive Modulation and Coding). | Request Band Width(bps) | Admitted BW(sps) | Number Of Admitted Connections | Number Of rejected Connections |
|---|---|---|---|---|
| QPSK-1/2 | 6128000 | 5152000 | 8 | 1 |
| 64 QAM | 11128000 | 2262800 | 9 | 0 |
| 16 QAM | 11128000 | 5087800 | 9 | 0 |

In above result show that in 16 QAM is better choice on proposed network. In both case we also observed that the connection has been rejected with Uplink i.e. in the duration of demand from SS to BS and the mostly service class GOLD and scheduling type UGS is rejected.

*3.2 WiMAX*

In this case we have observed the performance analysis of WiMAX network is done on the basis of Location Based Radio Resource Allocation (LBRRA). The Resources are allocated on the account of total Uplink (UL) capacity, total Downlink (DL) capacity, number of Admitted Connections and number of Rejected Connections for a particular Base Station (BS).

At BS1 and BS2 in the network of, the Resources which are allocated are shown in Table 10 and Table 11.

**Table 10**. Resource allocation at BS1

| Statistic | Value |
| --- | --- |
| Total Capacity (Msps) | 12.7104 |
| -> Total Uplink Capacity (Msps) | 2.6304 |
| -> Total Downlink Capacity (Msps) | 10.08 |
| Admitted Capacity (Msps) | 12.6424 |
| Number of Admitted Connections | 62 |
| Number of Rejected Connections | 19 |

**Table 11.** Resource allocation at BS2

| Statistic | Value |
| --- | --- |
| Total Capacity (Msps) | 12.7104 |
| -> Total Uplink Capacity (Msps) | 2.6304 |
| -> Total Downlink Capacity (Msps) | 10.08 |
| Admitted Capacity (Msps) | 12.6136 |
| Number of Admitted Connections | 44 |
| Number of Rejected Connections | 10 |

**4. Based on Proposed Algorithm**

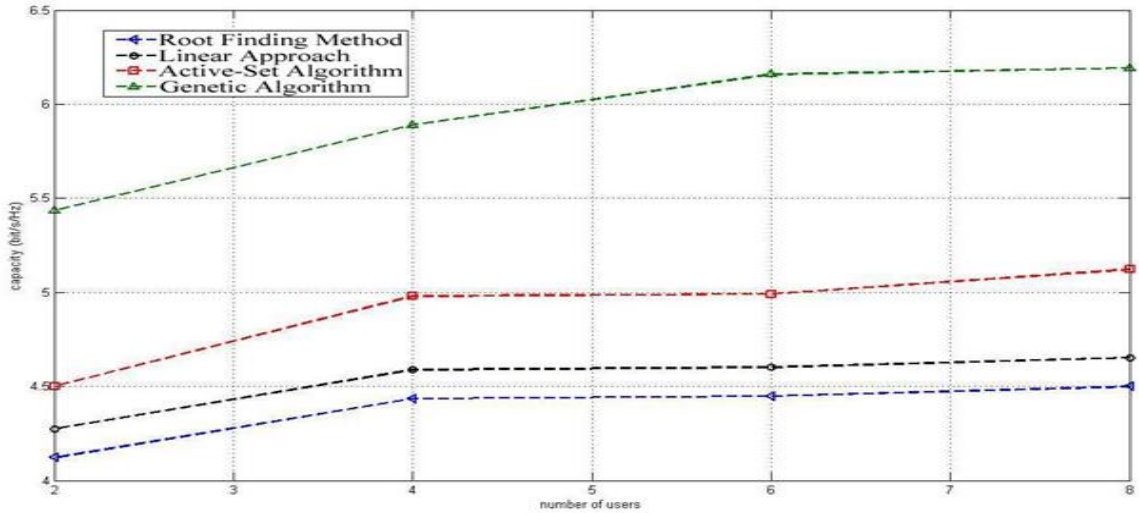

**Figure 2**. Total capacity versus number of users in downlink WiMAX network with N=64 and Subchannel SNR=50dB. The capacity achieved by the proposed Linear method , Active-Set and Genetic Algorithm Optimization technique for 1 to 8 subscribers.

Figure 2 shows the comparison of total capacities between the Proposed methods Linear , Active-Set and Genetic Algorithm. Notice that the capacities increase as the number of users increases. This is the effect of multiuser diversity gain, which is more prominent in WiMAX systems with larger number of users. The proposed methods have consistently higher total capacity than obsolete method like ROOT-FINDING method for all the numbers of users for this set of Simulation parameters. It is observed through simulation that the system using Genetic Algorithm performs better than Active-Set, Linear approach and ROOT-FINDING algorithms in terms to achieve highest capacity while being applicable to a more general class of systems.

## 5. Conclusion

The work described in this paper is based on LBRRA with two case studies, in first phase an exact analysis has been given for location based resource allocation for WiMAX network. Consequence has given apparent idea about the performance with different application with QoS in terms of how many connections have been rejected for designed or for proposed model. We have accomplished that mostly the connection has been rejected at all the Base Stations (BSs) in Uplink (UL) with the service class of GOLD with scheduling type of UGS. These rejected connections are happen due to ARQ parameters on UGS scheduling scheme. Automatic Retransmission Request (ARQ) can be enabled only on rtPS, nrtPS and BE flows. In second phase a literal analysis has been given for location based resource allocation for WiMAX and WLAN interface network. On the basis of simulation result we have accomplished that Total

Capacity of with UL and DL is constant. At this juncture also concluded that poling overhead is also constant and it is appeared due to uplink (UL) real time service. In last phase we then conciliated that in our location based proposed network the Request BW (Msps) and Admitted BW is different in all three cases with AMC and 16 QAM is best one with proposed location network.

An exact analysis has been given for location based resource allocation for WiMAX network. Result has given clear idea about the performance with different application with QoS in terms of how many connections have been rejected for designed or for proposed model. We have concluded that mostly the connection has been rejected at all the Base Stations (BSs) in Uplink (UL) with the service class of GOLD with scheduling type of UGS. These rejected connections are happen due to ARQ parameters on UGS scheduling scheme. Automatic Retransmission Request (ARQ) can be enabled only on rtPS, nrtPS and BE flows.

We recommended Active-Set algorithm for WiMAX systems because its consistency and low complexity. Latency period plays a vital role in real time applications handled by WiMAX systems. Base station should decide and assign priorities and resources within framed time duration, Active-Set algorithm is the best choice instead of using Genetic Algorithm or Linear approach. On the network if all subscribers are using same service like voice or data then we suggest using linear approach for allotting resources. In hybrid scenario Active- Set algorithm is worth. For off line decision making there is no option for Genetic Algorithm as we can compromise on latency period.

**Acknowledgment**

We would like to thank ECED department, SVNIT, Surat and OPNET Modeler.